\documentclass[12pt,a4paper]{article}
\usepackage{bm,graphicx}
\begin{document}
\textwidth=135mm
 \textheight=200mm
\begin{center}
{\bfseries Probing TMDs through azimuthal distributions of pions inside a jet in hadronic collisions
\footnote{{\small Talk given by C.P.\ at 20th International Symposium on Spin Physics (SPIN2012), JINR, 
Dubna (Russia), 17 - 22 September 2012.}}}
\vskip 5mm
U. D'Alesio$^{\dag,\ddag}$, F. Murgia$^\ddag$, C. Pisano$^\ddag$ 
\vskip 5mm
{\small {\it $^\dag$ Dipartimento di Fisica, Universit\`a di Cagliari, Cittadella Universitaria\\ I--09042 Monserrato (CA), Italy}} \\
{\small {\it $^\ddag$ INFN, Sezione di Cagliari, C.P. 170, I--09042 Monserrato (CA), Italy}}
\\
\end{center}
\vskip 5mm
\centerline{\bf Abstract}

The azimuthal distributions around the jet axis of leading pions produced in the jet fragmentation process in $pp$ collisions are studied within the framework of the so-called generalized parton model. The observable leading-twist azimuthal asymmetries are estimated in kinematic configurations presently investigated at RHIC. It is shown how the main contributions
coming from the Collins and Sivers effects can be disentangled. In addition, a test of the process dependence of the Sivers function is 
provided.

\vskip 10mm

The process $p^{\uparrow}p\to{\rm jet} \,\pi+X$, where one of the protons is in
a transverse spin state and the jet is produced with a large transverse 
momentum, $p_{{\rm j}T}$, is studied within the framework of the generalized 
parton model (GPM), in which factorization is assumed and spin and intrinsic 
parton motion effects are taken into account \cite{D'Alesio:2010am}. 
In this approach, azimuthal asymmetries in the distribution of leading pions 
around the jet axis are given by convolutions of different 
transverse momentum dependent distribution and fragmentation functions (TMDs). 
Similarly to the case of semi-inclusive deep inelastic scattering (SIDIS), 
it is possible to single out the different contributions by taking appropriate
 moments of such asymmetries. This would  be very useful in clarifying 
the role played mainly by the Sivers distribution and the Collins fragmentation 
function in the sizeable single spin
 asymmetries observed at RHIC for single inclusive pion production, where 
these underlying mechanisms cannot be disentangled.

\begin{figure}[t]
\begin{center}
 \includegraphics[angle=0,width=0.4\textwidth]{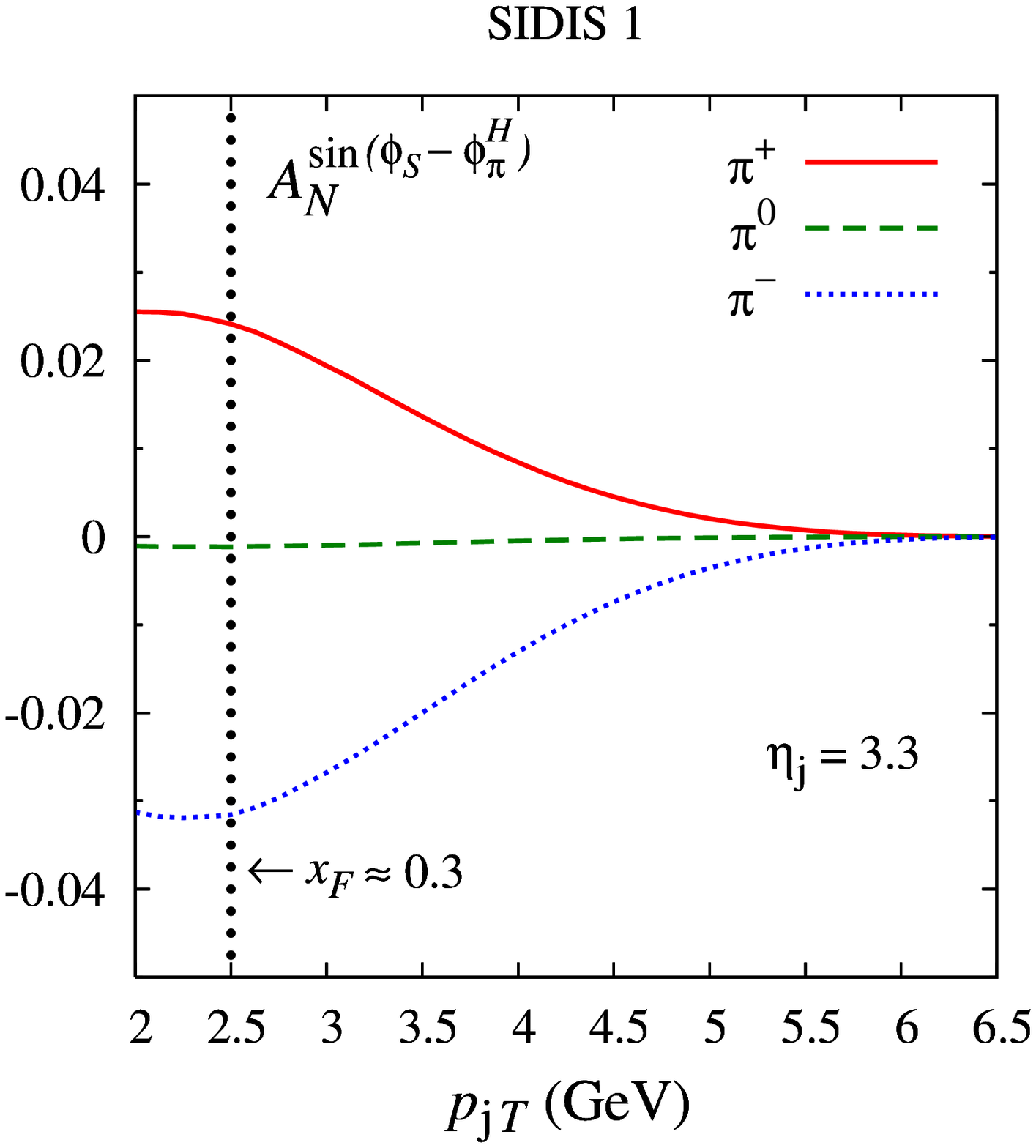}
 \includegraphics[angle=0,width=0.4\textwidth]{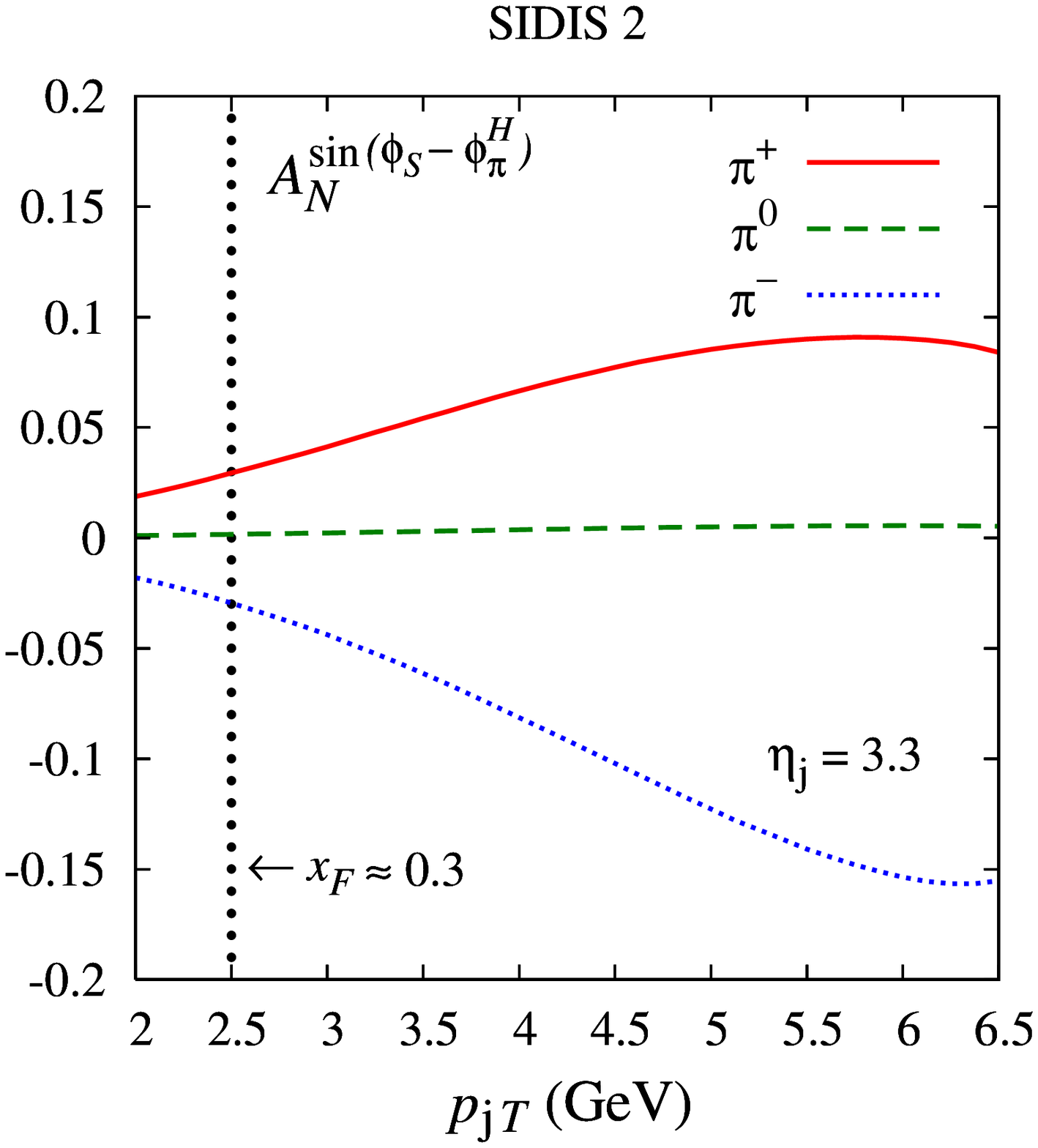}
 \caption{ The Collins asymmetry 
$A_N^{\sin(\phi_{S}-\phi_\pi^H)}$ as a function of $p_{{\rm j}T}$, at fixed jet rapidity $\eta_{\rm j} =3.3$
and energy   $\sqrt{s}=200$ GeV.
\label{asy-an-coll-par200} }
\end{center}
\end{figure}
\begin{figure}[t]
\begin{center}
 \includegraphics[angle=0,width=0.35\textwidth]{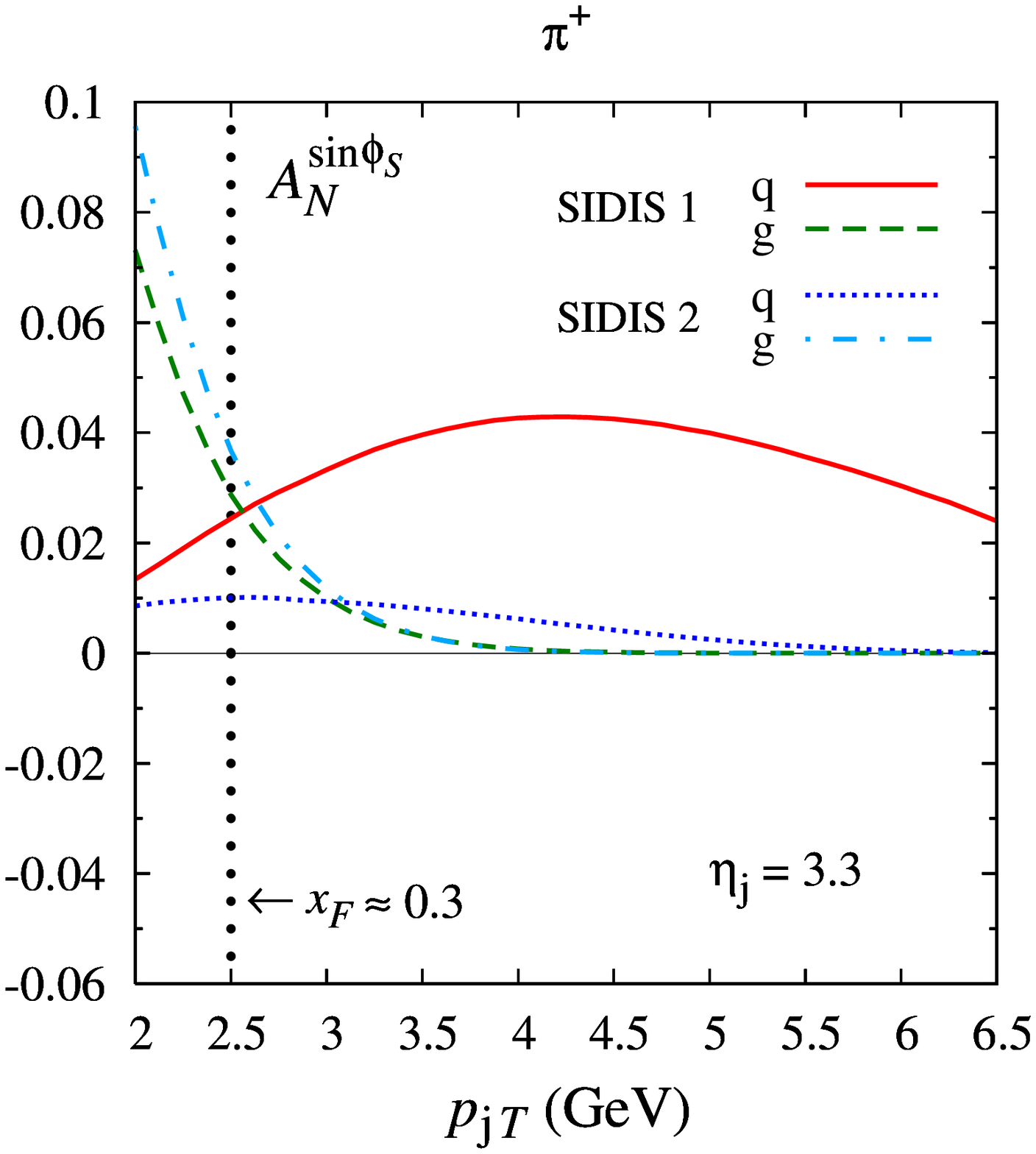}
 \hspace*{-20pt}
 \includegraphics[angle=0,width=0.35\textwidth]{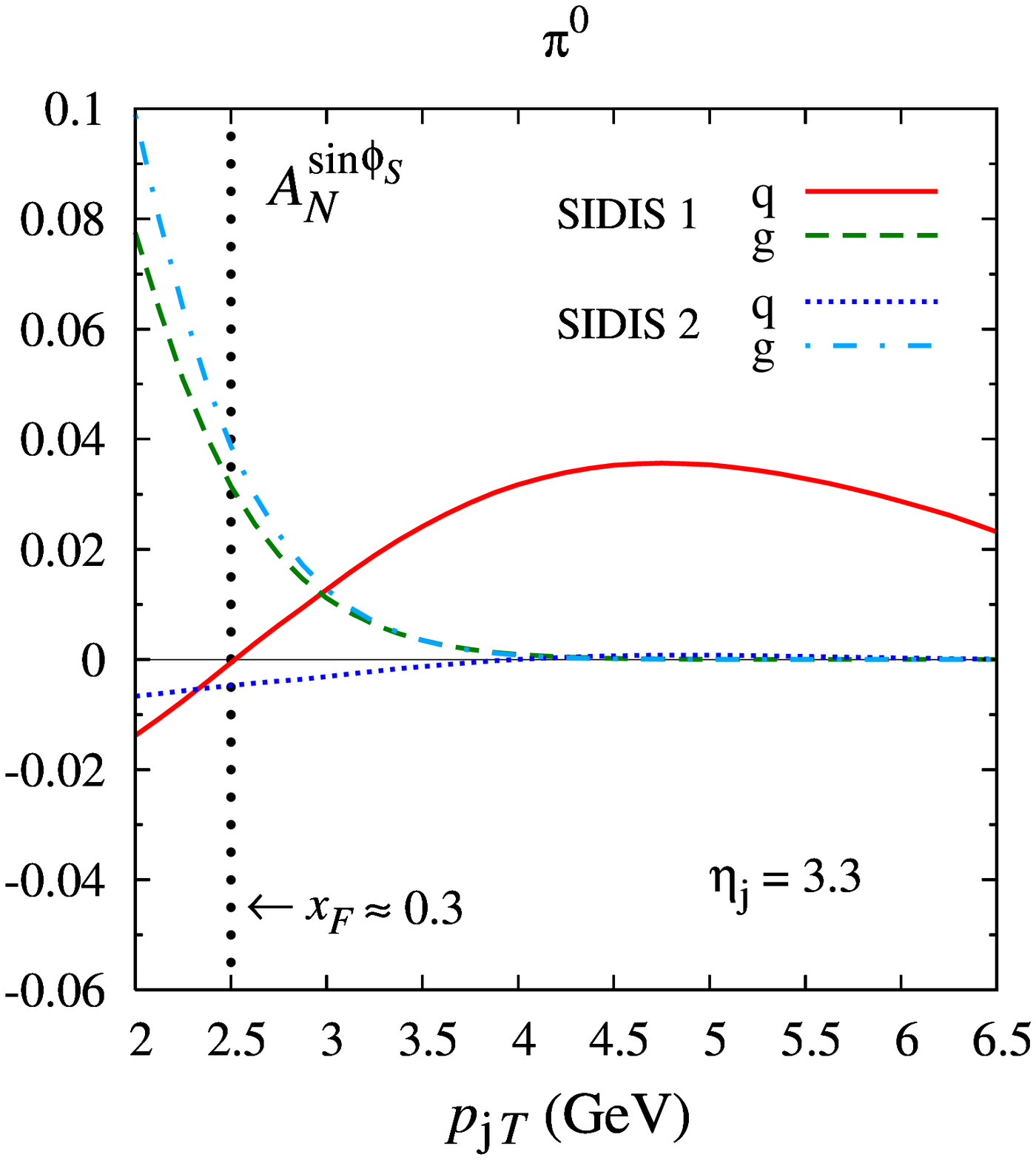}
 \hspace*{-20pt}
 \includegraphics[angle=0,width=0.35\textwidth]{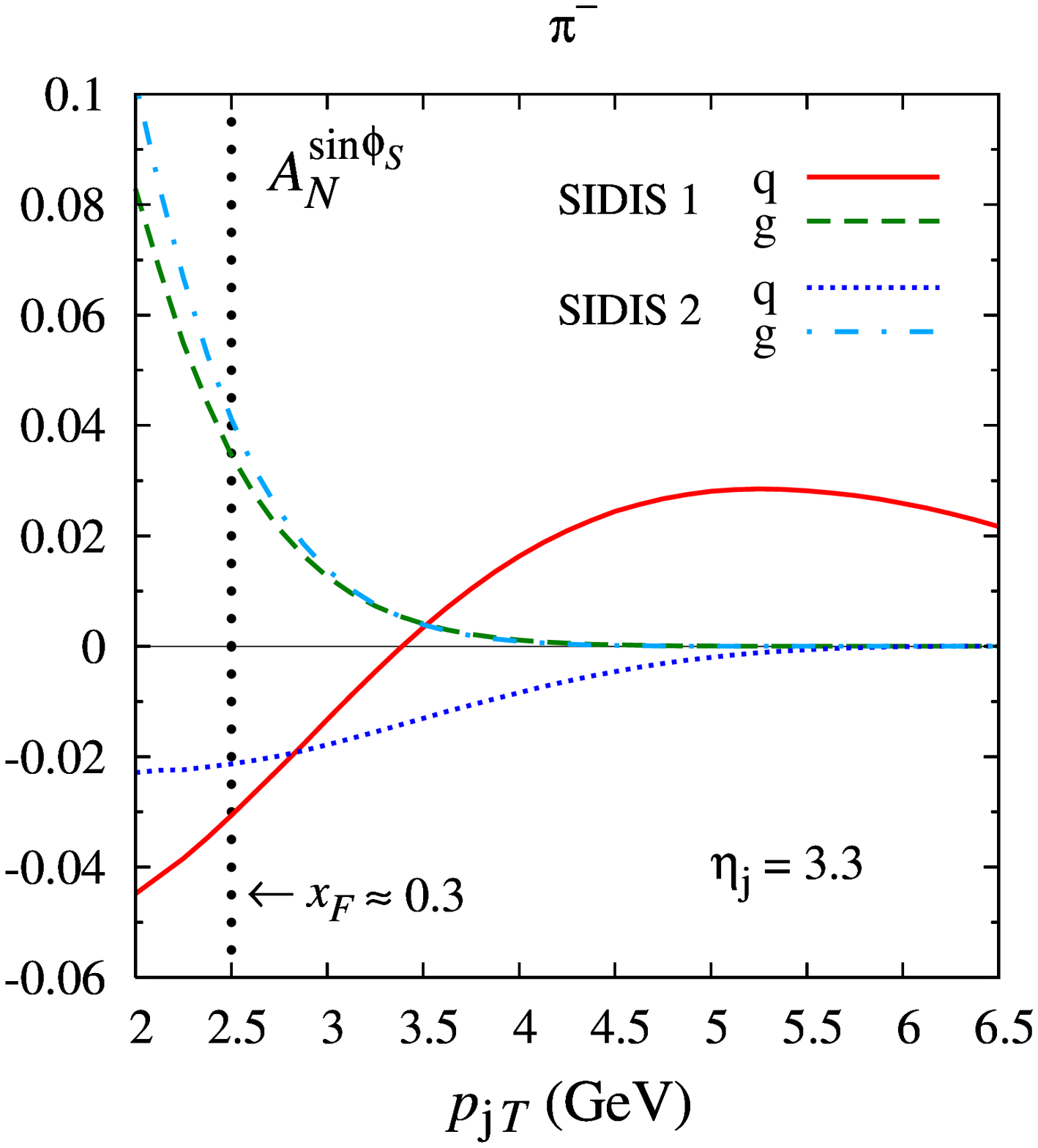}
 \caption{Same as in Fig.\ \ref{asy-an-coll-par200}  but for the Sivers asymmetry  $A_N^{\sin\phi_{S}}$.
 \label{asy-an-siv-par200} }
\end{center}
\end{figure}

The single-transverse polarized cross section for the process under study has 
been calculated at leading order in pQCD utilizing the helicity formalism  and 
has the general structure \cite{D'Alesio:2010am}
\begin{eqnarray}
2{\rm d}\sigma(\phi_{S},\phi_\pi^H)\!\! &\sim & \!\!{\rm d}\sigma_0
\, +\, {\rm d}\Delta\sigma_0\sin\phi_{S} \, +\, 
{\rm d}\sigma_1\cos\phi_\pi^H \, +\,  
{\rm d}\sigma_2\cos2\phi_\pi^H \nonumber \\
&& \quad +  \, {\rm d}\Delta\sigma_{1}^{-}\sin(\phi_{S}-\phi_\pi^H)\, +\, 
{\rm d}\Delta\sigma_{1}^{+}\sin(\phi_{S}+\phi_\pi^H)\nonumber \\
&& \quad\quad+\, {\rm d}\Delta\sigma_{2}^{-}\sin(\phi_{S}-2\phi_\pi^H) 
+ \, {\rm d}\Delta\sigma_{2}^{+}\sin(\phi_{S}+2\phi_\pi^H)\,,
\label{d-sig-phi-SA}
\end{eqnarray}
where  $\phi_S$ is the angle of the proton transverse spin vector $S$ relative 
to the jet production plane, and $\phi_\pi^H$ is the 
azimuthal angle of the pion three-momentum around the jet axis, as measured in the fragmenting parton helicity frame \cite{D'Alesio:2010am}. 
The various angular modulations can be projected out by defining the azimuthal 
moments
\begin{equation}
A_N^{W(\phi_{S},\phi_\pi^H)}
=
2\,\frac{\int{\rm d}\phi_{S}{\rm d}\phi_\pi^H\,
W(\phi_{S},\phi_\pi^H)\,[{\rm d}\sigma(\phi_{S},\phi_\pi^H)-
{\rm d}\sigma(\phi_{S}+\pi,\phi_\pi^H)]}
{\int{\rm d}\phi_{S}{\rm d}\phi_\pi^H\,
[{\rm d}\sigma(\phi_{S},\phi_\pi^H)+
{\rm d}\sigma(\phi_{S}+\pi,\phi_\pi^H)]}\,,
\label{gen-mom}
\end{equation}
with $W(\phi_{S},\phi_\pi^H)$ being one of the circular functions of $\phi_{S}$ and $\phi_\pi^H$  in (\ref{d-sig-phi-SA}). 

The upper bounds of all these different asymmetries have been evaluated for RHIC kinematics and can be found in  \cite{D'Alesio:2010am}. In the following only those (sizeable) effects are considered, that involve TMDs for which parameterizations are available from independent fits to SIDIS, Drell-Yan (DY), 
and $e^+e^-$ data.
The asymmetry $A_N^{\sin(\phi_{S}-\phi_\pi^H)}$ is given mainly by the convolution of the transversity distribution and the Collins fragmentation functions. 
It is shown in Fig.~\ref{asy-an-coll-par200} in the forward 
rapidity region adopting 
two different sets of parameterizations (SIDIS~1 and  SIDIS~2) 
\cite{D'Alesio:2010am}. Preliminary RHIC data 
\cite{Poljak:2011vu} are in agreement with our prediction of 
an almost vanishing  Collins asymmetry for neutral pions.
The quark and gluon contributions to the Sivers asymmetry $A_N^{\sin\phi_{S}}$,
which cannot be disentangled, are presented in Fig.~\ref{asy-an-siv-par200} in 
the same kinematic region. The quark term is obtained utilizing again the SIDIS~1 and SIDIS~2 parameterizations, while  the gluon Sivers function is tentatively taken positive and saturated to a bound obtained by considering PHENIX data for  inclusive neutral pion production at mid-rapidity \cite{D'Alesio:2010am}.
In both figures, the two parameterizations  
give comparable results only for values of the Feynman variable $x_F$ smaller 
than $0.3$, marked by the dotted vertical lines. Above this 
limit TMDs are not constrained by
 present SIDIS data, hence our predictions are affected by large uncertainties.
 A measurement of these asymmetries would therefore provide very useful 
information on the large $x$ behaviour of the underlying TMDs.

\begin{figure}[t]
\begin{center}
\includegraphics[angle=0,width=0.35\textwidth]{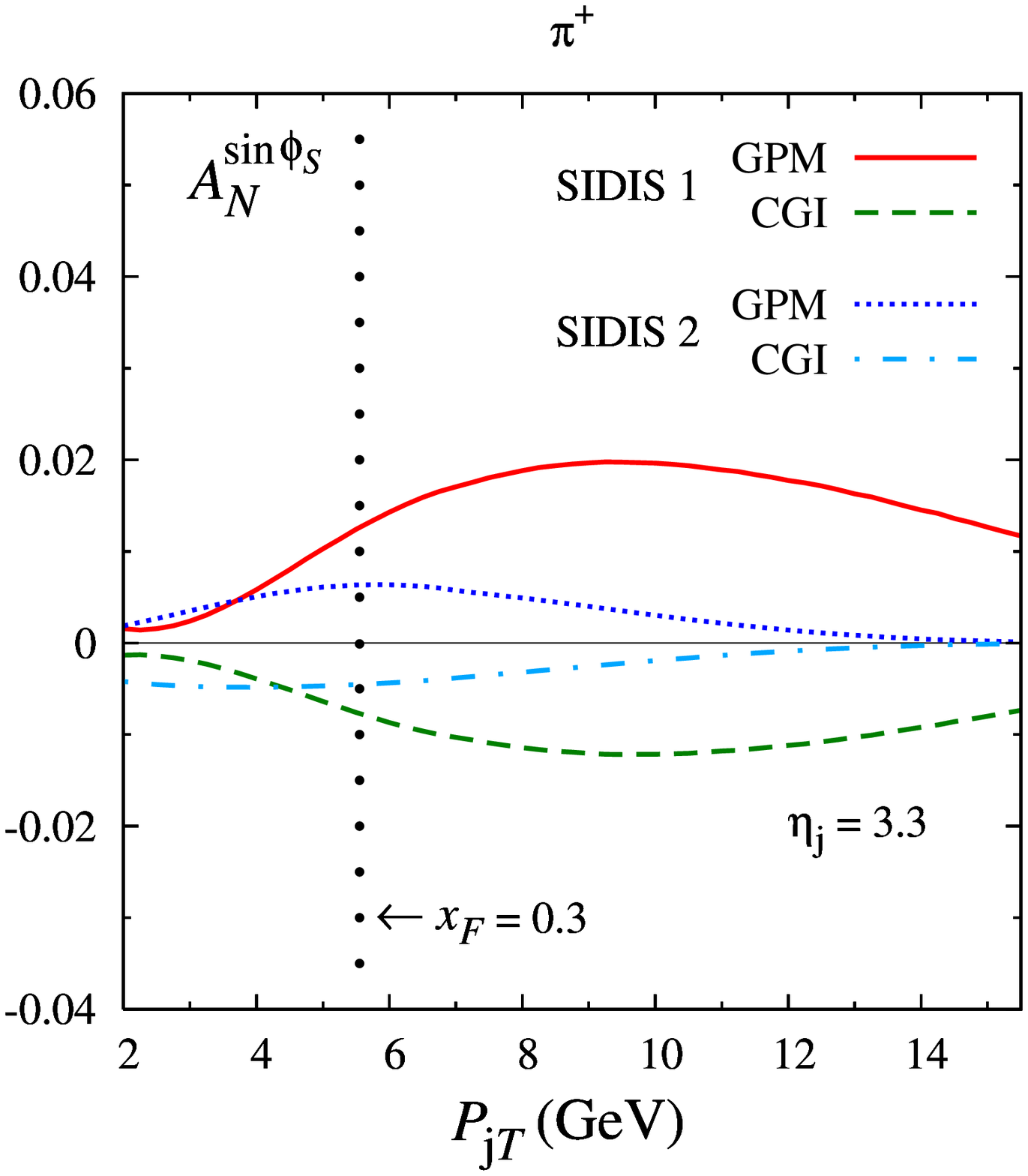}
 \hspace*{-20pt}
 \includegraphics[angle=0,width=0.35\textwidth]{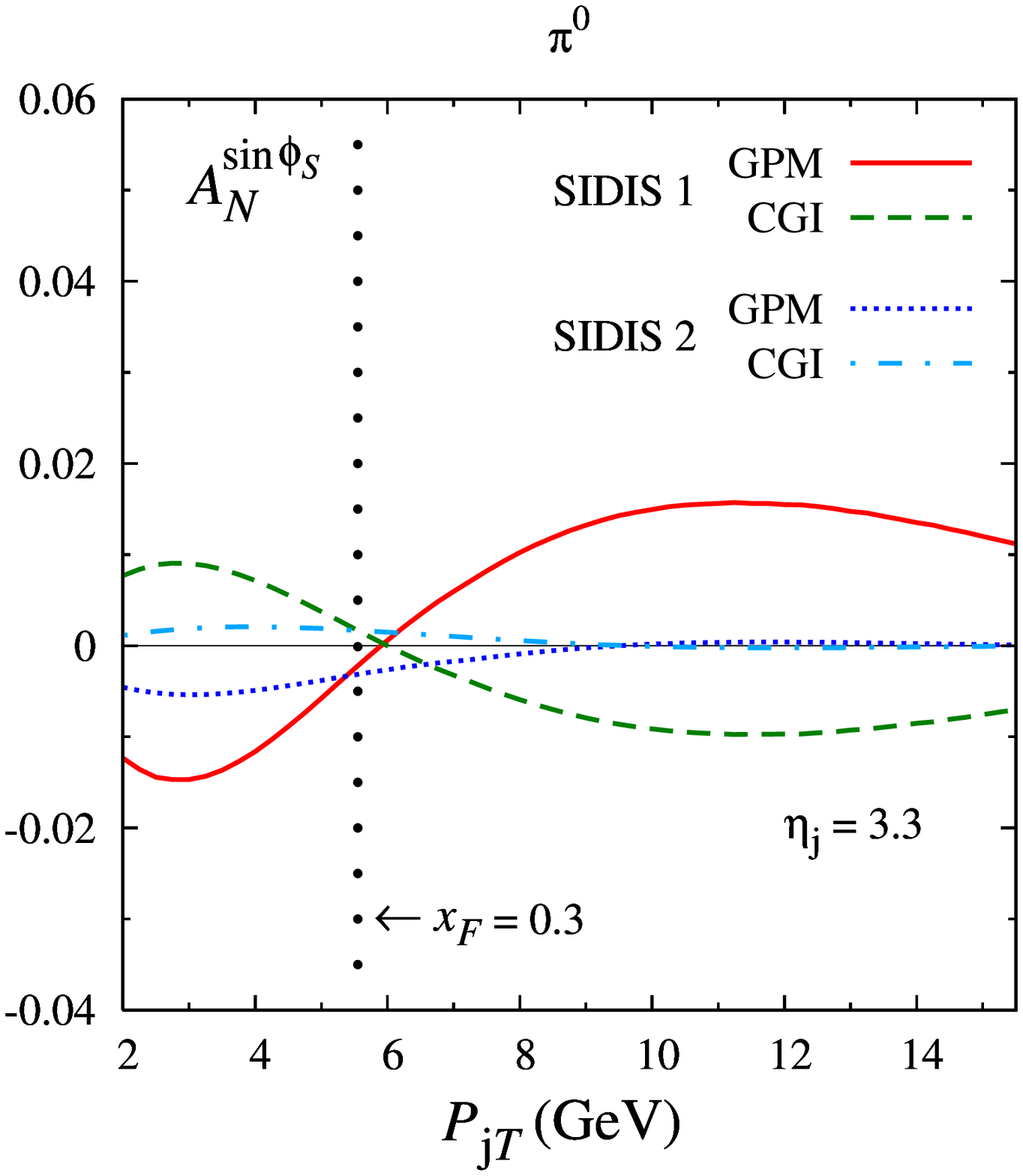}
 \hspace*{-20pt}
 \includegraphics[angle=0,width=0.35\textwidth]{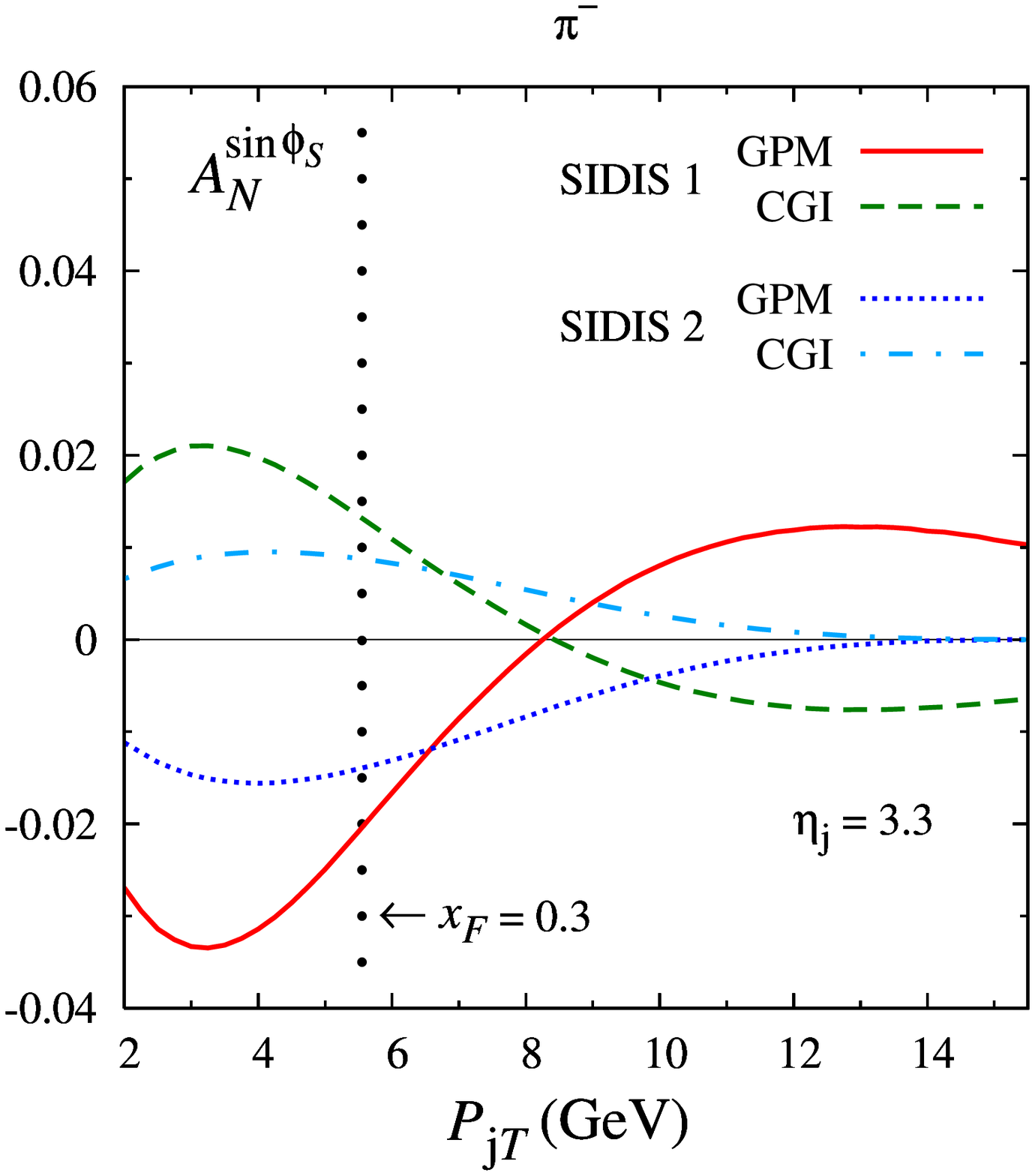}
\caption{The estimated quark contribution to the Sivers asymmetry $A_N^{\sin\phi_{S}}$ as a function of $p_{{\rm j}T}$, at fixed jet rapidity $\eta_{\rm j} =3.3$ 
 and energy $\sqrt{s}=500$ GeV.}
\label{fig3}
\end{center}
\end{figure}

So far TMDs have been  assumed to be universal. In the  framework 
of the color gauge invariant (CGI) GPM \cite{D'Alesio:2011mc}, one takes into account also the effects of initial (ISI) and final state interactions (FSI) 
between the active parton and the spectator remnants, which 
can render the TMDs process dependent. For example,  the Sivers functions in SIDIS and DY are expected to have opposite relative signs, due to the difference between FSI and ISI occurring, separately, in the two reactions. 
This is  a decisive prediction (not yet confirmed by experiments) of our 
present understanding of single spin asymmetries. The quark Sivers function turns out to have a more complicated color factor structure in $p^{\uparrow}p\to{\rm jet}\, \pi+X$, because
 both ISI and FSI contribute \cite{D'Alesio:2011mc}. Nonetheless, in the 
forward rapidity region only the $qg\to qg$ channel dominates. Therefore,
as shown in Fig.~\ref{fig3}, our 
results for the Sivers asymmetries obtained with and without the inclusion of
ISI and FSI have comparable sizes but opposite signs, 
in strong analogy with the DY case.  Hence the observation of a sizeable asymmetry could easily discriminate among the two approaches and test the process
dependence of the Sivers function. 

To conclude, single-spin asymmetries for inclusive jet production, 
described only by the Sivers function, have also been analysed 
\cite{D'Alesio:2010am,D'Alesio:2011mc}. The results obtained for 
$A_N^{\sin\phi_{S}}$  look very similar to the ones for jet-neutral pion 
production presented in the central panel of Fig.~\ref{fig3}. 
According to preliminary data, the Sivers asymmetries for these 
two  processes are small and positive
\cite{Poljak:2011vu,Nogach:2012}. Further comparison with experiments is 
needed to confirm the validity of the factorization assumption and test the universality properties of TMDs.

\vskip 10mm

We  acknowledge support from the FP7 EU-program HadronPhysics3 (Grant Agreement 283286). U.D.~and F.M.~acknowledge partial support by Italian MIUR (PRIN 2008).

\end{document}